# Mechanisms of Skyrmion and Skyrmion Crystal Formation from the Conical Phase ‡

Tae-Hoon Kim,[1]† Haijun Zhao,[1,2]† Ben Xu,[3] Brandt A. Jensen,[1] Alexander H. King,[1,4] Matthew J. Kramer,[1] Cewen Nan,[3] Liqin Ke,[1]* Lin Zhou[1]*

[1] Ames Laboratory, U.S. Department of Energy, Ames, Iowa 50011, USA

[2] School of Physics, Southeast University, Nanjing 211189, China

[3] School of Materials Science and Engineering, State Key Laboratory of New Ceramics and Fine Processing, Tsinghua University, Beijing 100084, China

[4] Department of Materials Science and Engineering, Iowa State University, Ames, IA 50011, USA

*Correspondence to: liqinke@ameslab.gov; linzhou@ameslab.gov

† These authors contributed equally to this work.

‡ Results of this paper was presented on Aug. 6, 2019, at Microscopy & Microanalysis 2019 (Portland, OR).

**Abstract**

Real-space topological magnetic structures such as skyrmions and merons are promising candidates for information storage and transport. However, the microscopic mechanisms that control their formation and evolution are still not clear. Here, using in-situ Lorentz transmission electron microscopy, we demonstrate that skyrmion crystals (SkXs) can nucleate, grow, and evolve from the conical phase in the same ways that real nanocrystals form from vapors or solutions. More intriguingly, individual skyrmions can also "reproduce" by division in a mitosis-like process that allows them to annihilate SkX lattice imperfections, which is not available to crystals made of mass-conserving particles. Combined string method and micromagnetic calculations show that competition between repulsive and attractive interactions between skyrmions governs particle-like SkX growth, but non-conservative SkX growth appears to be defect-mediated. Our results provide insights towards manipulating magnetic topological states by applying established crystal growth theory, adapted to account for the new process of skyrmion mitosis.

Crystal formation mechanisms play critical roles in controlling materials microstructures. Advances in *in situ* microscopy techniques have contributed significantly to our understanding of the interplay between the thermodynamics and kinetics of the processes of crystallization and crystal growth[1–5]. It is now widely accepted that the growth of a crystal proceeds either through monomer-by-monomer addition (MA) or by particle attachment (PA), or both mechanisms concurrently[2]. The formation of quasiparticle assemblies, such as skyrmion crystals (SkXs), presents new opportunities to explore phenomena that are not reflected in the growth of crystals made of real particles such as atoms or molecules.

A skyrmion is a nanoscale vortex-like spin structure with topological charge[6–10], and magnetic skyrmions are promising information carriers for spintronic applications[11]. Skyrmions are often observed in non-centrosymmetric magnetic materials with helical ground states, resulting from the competition between the exchange and Dzyaloshinskii-Moriya interactions (DMI)[12,13]. An external magnetic field perpendicular to the wavevector of the helical structure may induce a conical state, followed by formation of SkXs[13,14]. Skyrmions and SkXs form at thermodynamic equilibrium, depending on external magnetic fields, electric currents and temperature[6,8,13,15–19]. The equivalents of crystal lattice defects such as grain boundaries (GBs) and dislocations have been observed in SkXs[20–22].

Despite extensive imaging of SkXs[8,10,23] and a variety of proposed mechanisms for individual skyrmion formation[7,24,25], real-space observation of skyrmion winding from matrix phases remains challenging, due to the narrow range of temperature and external magnetic field over which the transition occurs. In this study, we have achieved real-time monitoring of SkX nucleation in β-Mn-type $Co_8Zn_{10}Mn_2$ thin films using Lorentz transmission electron microscopy (LTEM). We discover that the formation of SkX includes two processes that overlap in time. The first is nucleation of skyrmions by coordination of spin, which proceeds through an intermediate state. The second is growth of SkXs, which follows a nucleation and growth mechanism of typical first-order phase transformation, by the addition of subsequent skyrmions. SkXs grow through the addition of individual skyrmions on their peripheries, the addition of skyrmion clusters, and by a new non-conservative growth mechanism of internal skyrmion splitting. Micromagnetic simulation together with string method reveals that the inter-skyrmion attractive forces promote the formation of skyrmion clusters, while the skyrmion self-splitting process is due to anisotropic stretching forces from its miscoordinated neighbors.

**Results and Discussion**

**Nucleation of skyrmion and SkX**

Nucleation of the first skyrmion is a crucial step toward SkX formation and represents transformation from the conical phase with a corresponding change of topological number from zero to one. It is also the most challenging process to monitor. We have overcome the experimental difficulty by slowing down the kinetics using precisely controlled temperature, magnetic field, and a rapid camera speed of 50 ms per frame. First, we systematically examined the temperature and external magnetic field dependence of different thermal-equilibrium states. A phase diagram with a contour plot of skyrmion phase density based on LTEM observation (fig. S1) shows that skyrmions grow from the conical phase above 80 ℃ at an external field higher than 100 mT in a ~200 nm thick (110) plate (fig. S2). Accordingly, the sample was heated up quickly to 84 ℃ under 135 mT so that the conical-to-skyrmion transition proceeded through a first-order process[26]. The helical phase observed at room temperature under zero field (Fig. 1A) transformed into a single conical phase upon reaching the targeted experimental condition (Fig. 1B). A skyrmion embryo, with about 60% of the contrast of a fully-formed skyrmion, appeared after 1.85 s of incubation time (Fig. 1C), developing into a full-contrast skyrmion with a slight shift from the initial position at 2.6 s (Fig. 1D). Both the skyrmion embryo and the skyrmion were mobile, undergoing random movements akin to Brownian motion.

The region around the first skyrmion then acted as a preferred nucleation site for a second skyrmion[26], and a two-skyrmion complex was formed with a specific separation after 5.75 s (Fig. 1E). Subsequent skyrmions grew at a faster rate (Figs. 1F-H), possibly due to formation of more skyrmion-skyrmion bonds. Skyrmion trimer forms with a triangular structure and skyrmion tetramer takes the shape of a rhombus. The oligomers grow to clusters with further addition of skyrmions. As with the first skyrmion, contrast evolution was observed for the third and fourth ones, as shown in intensity profile comparison (Fig. 1O, see fig. S3 for time-resolved evolution). Because all Lorentz images were taken under the same defocus condition, the contrast transformation of the same "particle" is approximately proportional to the projected magnetization change along the electron beam direction[27], suggesting the formation of intermediate states, during nucleation of skyrmion, similar to skyrmion lattice decay[28].

**Propagation of SkX through MA and PA mechanisms**

When there were more than two skyrmions in the cluster, SkX lattice is observed to rotate and translate, along with the formation and addition of more skyrmions (Figs. 1H-K). The growth of the SkX cluster results in pronounced facet development (Fig. 1M-N), corresponding to {100} of the hexagonal system. The initial skyrmion cluster develops by the formation of single skyrmions at or very close to their coordination sites, in what is essentially a MA mechanism (Fig. 1D-N), as described in classical crystal growth theory. The lattice parameter, i.e., the core to core distance between adjacent skyrmions, was almost constant ($138 \pm 5$ nm) after adding additional skyrmions. Movie S1 shows the nucleation and growth of this cluster.

"Particle" attachment events were observed when two growing skyrmion clusters were close to each other. The merging of two clusters typically includes two processes: connection and lattice reorientation. Depending on the cluster size, they occur either concurrently or sequentially (movie S2). When both clusters are small, the skyrmion lattices rotate rapidly, and they rotate to a small misorientation during attachment (Figs. 2A-C). The misorientation angle between two clusters reduced from 26° to 3.5° within 50 ms when the slightly smaller cluster jumped into the larger one which was initially ~400 nm away (Figs. 2B-C). If a small cluster merges into a much larger one, attachment occurs first, forming a grain boundary in the SkX, followed by reorientation of the smaller grain. In the case shown in fig. S4, the initial misorientation angle of 21° was reduced to 6°, 2.1 s after the attachment. These types of cluster attachment are very similar to nanoparticle agglomeration, *via* the so-called oriented attachment mechanism, which includes three-dimensional rotation to reduce interparticle misalignment, and atom-by-atom reorientation after attachment through dislocation or grain boundary migration[3,29,30]. When both skyrmion grains are large and immobile, merging typically starts with the formation of a bridge between them, followed by skyrmion growth and lattice rearrangement (movie S3). As shown in Figs. 2D-E, a bridge with two skyrmion lattice planes was first created when the grains were ~300 nm apart. Subsequently, skyrmions grew around the high-curvature regions of the bridge than growing near the flat facet (Fig. 2F)[3]. At the same time, SkX rearrangement was accompanied, and slight misorientation reduction (from 7.8° to 5.4°) was observed. SkX thus also grows through the PA pathway[2].

**Dislocation annihilation through self-splitting mechanism**

Lattice defects can be formed either at the surface or inside the SkX. Surface kinks with fivefold coordination provide preferred sites for MA skyrmion growth (fig. S5, movie S4), which is the major pathway for SkX propagation. Edge dislocations, formed either at GB by the merger of SkX particles or by structural relaxation, are geometrically observed as adjacent fivefold and sevenfold-coordinated skyrmions (5-7 defect, Fig. 3A), which can be described equivalently as wedge disclination dipoles. This type of imperfect skyrmion coordination is able to evolve by a mechanism that is not possible in a crystal made of real molecules (movie S5): a skyrmion with seven neighbors (Fig. 3A) rapidly divides into two skyrmions (Fig. 3B). As a result, the position of the disclination dipole or edge dislocation shifts closer to the cluster edge, like a dislocation climbing by the attachment of an interstitial atom. The newly created skyrmion triggered local SkX lattice rearrangement (Fig. 3C), just as an edge dislocation climbing in a grain boundary toward the surface will reduce the misorientation of two crystals. In this case, however, the climb is facilitated by the spontaneous creation of a new crystal lattice quasi-particle rather than the addition or subtraction of an atom by diffusion from or to the surface.

**Structural relaxation**

SkX lattice rearrangement continues after skyrmions fully replace the conical phase. Splitting of the FFT spots was clearly visible right after the phase transition completed, implying the existence of differently oriented domains (fig. S6). Single domain SkX was formed 30s later, as confirmed by the disappearance of the satellite FFT spots (movie S6). This process proceeds through grain boundary defect reduction by internal skyrmion growth or rearrangement[20,22].

**Interaction force between asymmetric skyrmions**

The interaction force that controls crystallization of SkX from the conical phase is directly calculated by gradually bringing two well-separated skyrmions together using a combined micromagnetic and string method (supplementary 2.1-3), which is capable of searching the lowest transition path between two states. Fig. 4A shows the calculated energy profiles as a function of distance between two skyrmions separated in $x$ or $y$ direction with different sample thicknesses. All interaction profiles exhibit minima similar to the repulsive core and attractive tail Lennard-Jones functions. The attractive part is driven by reducing of cone-skyrmion interface energy. As shown in Fig. 4B, isolated skyrmions are twisted by the conical phase along z direction, which results in high cone-skyrmion interfacial energy[26]. When two skyrmions touch and form a skyrmion-skyrmion interface, both of them are less twisted. At the same time, the cone-skyrmion interface energy is minimized (Figs. 4C-E). Note that, due to the spiral nature of the conical phase, the interaction is orientation dependent as well. Thus, a slight difference between the energy profiles along the $x$ and $y$ direction is observed when the film thickness decreases from $3.8L_D$ to $1.9L_D$, where $L_D$ is the periodicity of the helical phase. Particles or quasi-particles in 2D systems with this kind of interaction favor clusters with hexagonal coordination[31], similar to ferrofluid, colloidal systems, and vortices in $MgB_2$ or low-κ superconductor[32], except that the skyrmion system belongs to the hardcore interaction. Similarly, merging of separated clusters reduced cone/skyrmion interfacial energy and promotes cluster attachment, as observed in Fig. 2.

**Micromagnetic simulation of self-splitting process**

To study self-splitting process similar to our experiment, we use a combined micromagnetic and string method to search for the lowest energy transition path for forming a new skyrmion at 5-7 defect. Our calculation shows that SS is the energetically favored path. We create an initial state with a cluster containing 5-7 defect (Fig. 5B), mimic our experiment. Note that, the sevenfold-coordinated skyrmions is elongated due to anisotropic stretching forces from its miscoordinated neighbors. We then replace this skyrmion with two skyrmions and relax the system to get another cluster, which is the target state after a new skyrmion is added. After that, we use our string method to search for the lowest energy transition pathway between these two clusters.

Fig. 5A shows the evolution of energy and skyrmion number $N_s = \frac{1}{4\pi} \int \int \boldsymbol{m} \cdot (\frac{\partial \boldsymbol{m}}{\partial x} \times \frac{\partial \boldsymbol{m}}{\partial y}) dx dy$, where **m** is a unit vector parallel to magnetization direction during SS. Representative spin configurations during the splitting process are shown in Fig. 5B-D. The SS starts at the surface of the sample by creating a magnetic monopole (Fig. 5D). The monopole then moves toward the opposite surface, and unzip the skyrmion into two (Figs. 5E-F). Our result agrees with Ref.[33], in which the opposite process, skyrmion merging was studied (details in supplementary 2.4).

**Conclusions**

In summary, the nucleation and growth of skyrmion and SkX from the conical phase in β-Mn-type $Co_8Zn_{10}Mn_2$ thin films are directly observed by precise field and temperature control. Skyrmions grow through intermediate states before penetrating through the whole film. Analogous to crystallization in molecular systems, both MA and PA mechanisms are clearly demonstrated. Moreover, skyrmion addition can occur inside a growing SkX by a self-splitting mechanism that has no analog in molecular crystals. Micromagnetic simulation combined with string method demonstrates that the crystal-like SkX formation is driven by a Lennard-Jones like skyrmion-skyrmion interaction forces. The self-spitting of a skyrmion at 5-7 defect is energetically favorable and proceed by creating a magnetic monopole near the surface, and then unzips the skyrmion into two. Our discovery provides an essential step toward understanding and manipulating the evolution of magnetic topological states for applications in quantum information technologies.

**Materials and Methods**

The bulk $Co_8Zn_{10}Mn_2$ sample with critical temperature $T_c \sim 370$ K was prepared by first sealing individual metals (all > 99.9% metals basis) in a quartz ampoule backfilled with ultra-high purity argon. The ampoule was placed into a furnace and heated to 1000 °C for 12 hours, then cooled at 1 °C/hr to 925 °C and held for 96 hours before quenching into water. Magnetic measurements were performed using a Quantum Design VersaLab™ vibrating sample magnetometer. A thin, polycrystalline piece was polished so the sample dimensions were greater than 5:1 aspect ratio. The sample was field cooled at a rate of 2 K/min under an applied field of H = 20 Oe. The polycrystalline sample shows a Curie temperature of 362 K and magnetization of 0.17 $\mu B$/f.u. under an applied field of H = 20 Oe.

A $Co_8Zn_{10}Mn_2$ (110) thin plate was fabricated using focused ion beam system (FIB, FEI Helios NanoLab G3). The crystal orientation was verified by electron backscattered diffraction analysis before lift-out. The plate was approximately 200 nm thick. The thickness map of the sample (fig. S2) was obtained by electron energy loss spectroscopy with Gatan Quantum ER 965.

*In-situ* Lorentz transmission electron microscopy (LTEM) observation was conducted on an FEI Titan Themis by using FEI NanoEx-Tm-i/v *in-situ* TEM holder which enabled rapid heating and precise temperature control under isothermal conditions. The external magnetic field was applied along the electron beam direction by partially exciting the objective lens. LTEM videos were taken at 20 fps by FEI Ceta camera to record skyrmion dynamics. The number of skyrmions was counted using Image J. The in-plane magnetization maps of the magnetic structure were obtained by the LTEM Fresnel images with a phase-retrieval QPt software on the basis of the transport of intensity equation[34].

**Supplementary Materials:**

Section 1. Supplementary figures and videos

Figures S1-S6, Movies S1-S6

Section 2. Simulation details

Figure S7, Movies S7-S9

# References

1. Martin, S. W. *et al.* Real-Space Imaging of Nucleation and Growth in Colloidal Crystallization. **292**, 258–262 (2001).

2. De Yoreo, J. J. *et al.* Crystallization by particle attachment in synthetic, biogenic, and geologic environments. *Science (80-. ).* **349**, aaa6760 (2015).

3. Li, D. *et al.* Direction-Specific Interactions. *Science (80-. ).* **336**, 1014–1018 (2012).

4. Liao, H. *et al.* Facet development during platinum nanocube growth. **345**, 916–919 (2014).

5. Chen, J. *et al.* Building two-dimensional materials one row at a time: Avoiding the nucleation barrier. **1139**, 1135–1139 (2018).

6. Mühlbauer, S. *et al.* Skyrmion lattice in a chiral magnet. *Science (80-. ).* **323**, 915–919 (2009).

7. Rößler, U. K., Bogdanov, A. N. & Pfleiderer, C. Spontaneous skyrmion ground states in magnetic metals. *Nature* **442**, 797–801 (2006).

8. Yu, X. *et al.* Aggregation and collapse dynamics of skyrmions in a non-equilibrium state. *Nat. Phys.* 1–5 (2018).

9. Nayak, A. K. *et al.* Magnetic antiskyrmions above room temperature in tetragonal Heusler materials. *Nature* **548**, 561–566 (2017).

10. Yu, X. Z. *et al.* Real-space observation of a two-dimensional skyrmion crystal. *Nature* **465**, 901–904 (2010).

11. Venema, L. *et al.* The quasiparticle zoo. *Nat. Phys.* **12**, 1085–1089 (2016).

12. Fert, A., Reyren, N. & Cros, V. Magnetic skyrmions: Advances in physics and potential applications. *Nat. Rev. Mater.* **2**, 17031 (2017).

13. Nagaosa, N. & Tokura, Y. Topological properties and dynamics of magnetic skyrmions. *Nat. Nanotechnol.* **8**, 899–911 (2013).

14. Tanigaki, T. *et al.* Real-Space Observation of Short-Period Cubic Lattice of Skyrmions in MnGe. *Nano Lett.* **15**, 5438–5442 (2015).

15. Karube, K. *et al.* Robust metastable skyrmions and their triangular-square lattice structural transition in a high-temperature chiral magnet. *Nat. Mater.* **15**, 1237–1242 (2016).

16. Fert, A., Cros, V. & Sampaio, J. Skyrmions on the track. *Nat. Nanotechnol.* **8**, 152–156 (2013).

17. Iwasaki, J., Mochizuki, M. & Nagaosa, N. Current-induced skyrmion dynamics in constricted geometries. *Nat. Nanotechnol.* **8**, 742–747 (2013).

18. Woo, S. *et al.* Observation of room-temperature magnetic skyrmions and their current-driven dynamics in ultrathin metallic ferromagnets. *Nat. Mater.* **15**, 501–506 (2016).

19. Oike, H. *et al.* Interplay between topological and thermodynamic stability in a metastable magnetic skyrmion lattice. *Nat. Phys.* **12**, 62–66 (2016).

20. Rajeswari, J. *et al.* Filming the formation and fluctuation of Skyrmion domains by cryo-Lorentz Transmission Electron Microscopy. **112**, 14212–14217 (2015).

21. Matsumoto, T. *et al.* Direct observation of S7 domain boundary core structure in magnetic skyrmion lattice. *Sci. Adv.* **2**, e1501280–e1501280 (2016).

22. Pöllath, S. *et al.* Dynamical Defects in Rotating Magnetic Skyrmion Lattices. *Phys. Rev. Lett.* **118**, 1–6 (2017).

23. Peng, L. *et al.* Relaxation Dynamics of Zero-Field Skyrmions over a Wide Temperature Range. *Nano Lett.* **18**, 7777–7783 (2018).

24. Heinze, S. *et al.* Spontaneous atomic-scale magnetic skyrmion lattice in two dimensions. *Nat. Phys.* **7**, 713–718 (2011).

25. Leonov, A. O. & Inoue, K. Homogeneous and heterogeneous nucleation of skyrmions in thin layers of cubic helimagnets. *Phys. Rev. B* **98**, 1–9 (2018).

26. Loudon, J. C., Leonov, A. O., Bogdanov, A. N., Hatnean, M. C. & Balakrishnan, G. Direct observation of attractive skyrmions and skyrmion clusters in the cubic helimagnet $Cu_2OSeO_3$. *Phys. Rev. B* **97**, 134403 (2018).

27. Chess, J. J. *et al.* Streamlined approach to mapping the magnetic induction of skyrmionic materials. *Ultramicroscopy* **177**, 78–83 (2017).

28. Wild, J. *et al.* Entropy-limited topological protection of skyrmions. *Sci. Adv.* **3**, 1–7 (2017).

29. Van Huis, M. A. *et al.* Low-temperature nanocrystal unification through rotations and relaxations probed by in situ transmission electron microscopy. *Nano Lett.* **8**, 3959–3963 (2008).

30. Penn, R. L. & Banfield, J. F. Imperfect oriented attachment: Dislocation generation in defect-free nanocrystals. *Science (80-. ).* **281**, 969–971 (1998).

31. Zhao, H. J., Misko, V. R. & Peeters, F. M. Analysis of pattern formation in systems with competing range interactions. *New J. Phys.* **14**, 063032 (2012).

32. Zhao, H. J., Misko, V. R., Tempere, J. & Nori, F. Pattern formation in vortex matter with pinning and frustrated intervortex interactions. *Phys. Rev. B* (2017). doi:10.1103/PhysRevB.95.104519

33. Milde, P. *et al.* Unwinding of a skyrmion lattice by magnetic monopoles. *Science (80-. ).* **340**, 1076–1080 (2013).

34. Kazuo Ishizuka (HREM Research Inc., Matsukazedai, S. Phase measurement in electron microscopy Using the Transport of Intensity Eqaution. 22–24 (2005).

**Acknowledgments: Funding**: This work is supported in part by Laboratory Directed Research and Development funds through Ames Laboratory (T.K, H.Z., L.Z.) and by the U.S. Department of Energy, Office of Science, Basic Energy Sciences, Materials Science and Engineering Division. Ames Laboratory is operated for the U.S. Department of Energy by Iowa State University under Contract No. DE-AC02-07CH11358. H.Z. is also supported in part by National Natural Science Foundation of China (No. 11704067). B.X. is supported by Major Program of National Natural Science Foundation of China (No. 51790494). L.K. was supported by the U.S. Department of Energy, Office of Science, Office of Basic Energy Sciences, Materials Sciences and Engineering Division, Early Career Research Program following conception and initial work supported by LDRD. All TEM and related work were performed using instruments in the Sensitive Instrument Facility in Ames Lab. **Author contributions**: L.Z. and L. K. conceived the project. T. K. carried out the microscopy work and analyzed the results together with L.Z.; H. Z. and B. X. carried out the micromagnetic calculation and analyzed the results together with L.K. and L.Z.; B. J. grew the crystal and performed magnetic property measurement. All authors contributed to the discussion of the results. L. Z., A. K. and L. K. wrote the manuscript with input from T. K., H. Z., B. J.. **Competing interests:** The authors declare no competing interests. **Data and materials availability:** All data are available in the manuscript or the supplementary material.

**Figures**

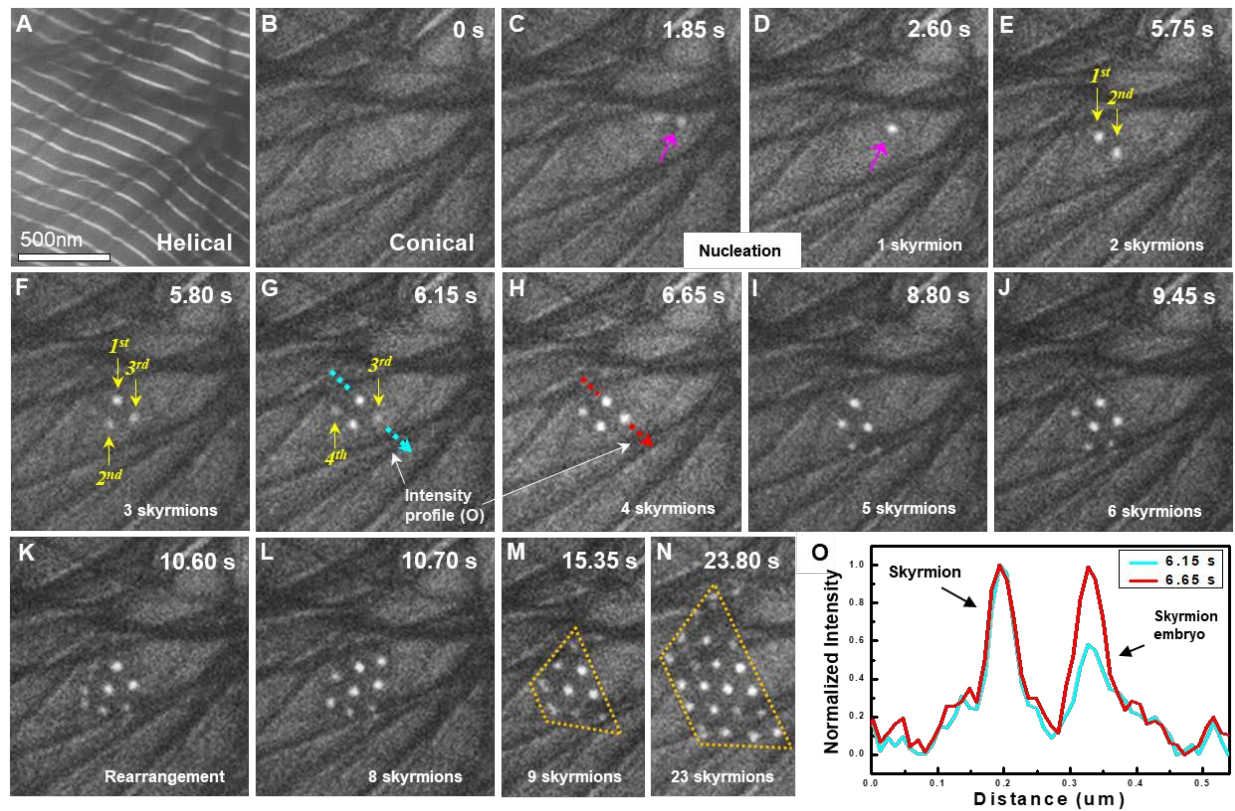

**Fig. 1 Skyrmion and SkX nucleation.** (**A**) Helical phase with periodicity of ~115 nm at room temperature under 0 mT. (**B-N**) Sequence of LTEM images from movie S1. Time indicates time progress after sample reached 84 ℃ under 135 mT. (**B**) Conical phase. (**C**) Skyrmion embryo with weak contrast. (**D**) Formation of an isolated skyrmion. Note the skyrmion position change from C to D. (**E-H**) SkX growth through MA with obvious intensity evolution of each skyrmion. (**I-L**) SkX growth accompanied with rearrangement. (**M, N**) Cluster of 9 and 23 skyrmions with facet. (**O**) Comparison of intensity profiles obtained from blue line in (**G**) and red line in (**H**).

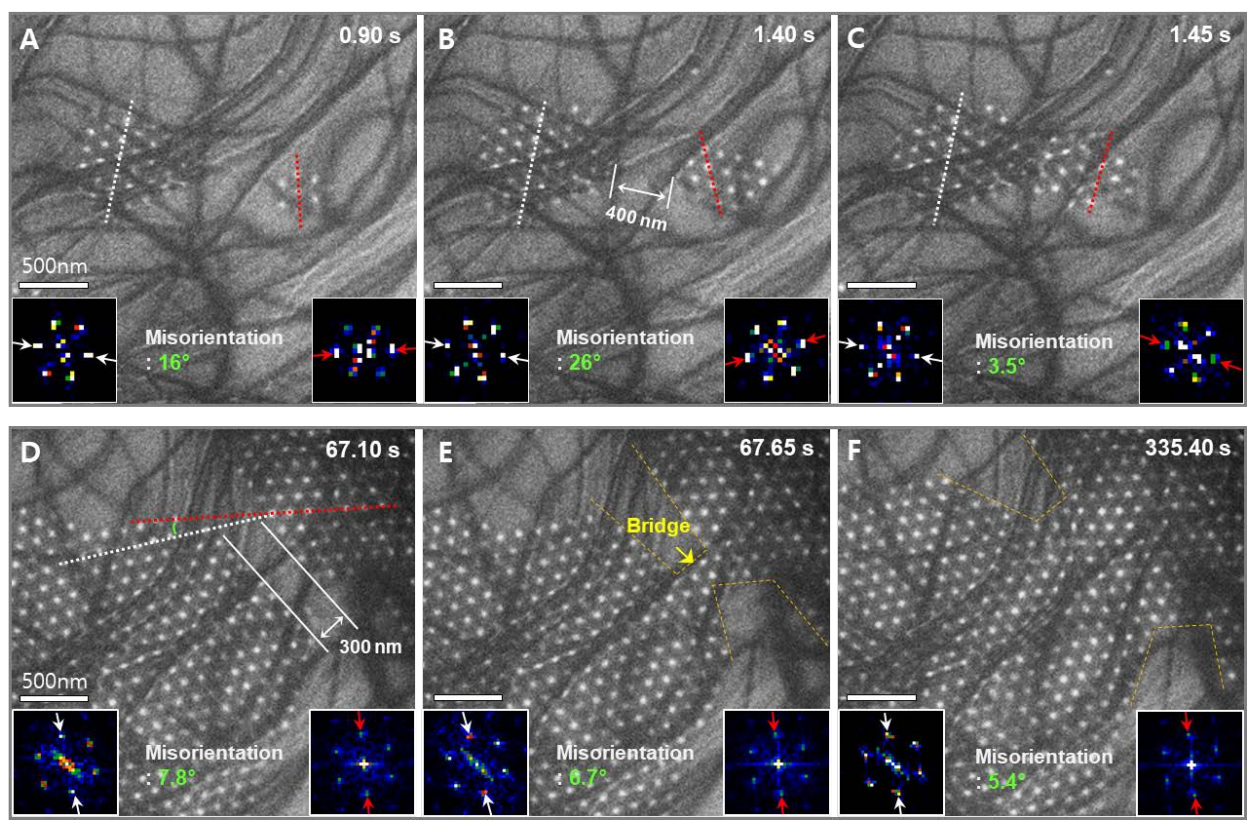

**Fig. 2 Coalescence by oriented attachment and relaxation of SkX. A-C** and **D-F** are sequences of LTEM images from movie S2 and S3, taken at 88℃ under 100 mT, respectively. **(A)** Two small SkX clusters. Insets show FFT obtained from each cluster. White and red arrows in FFT correspond to planes indicated by white and red lines in LTEM images. **(B)** Two clusters approached close to 400 nm and the misorientation angle changed to 26° by oscillation. **(C)** A small cluster jumped and merged into a larger cluster within 0.05 s and misorientation angle reduced to 3.5° simultaneously. **(D)** Two large clusters are ~300 nm apart. **(E)** Two clusters were attached by a bridge formation. **(F)** Skyrmions grow around the bridge. Misorientation was slightly reduced.

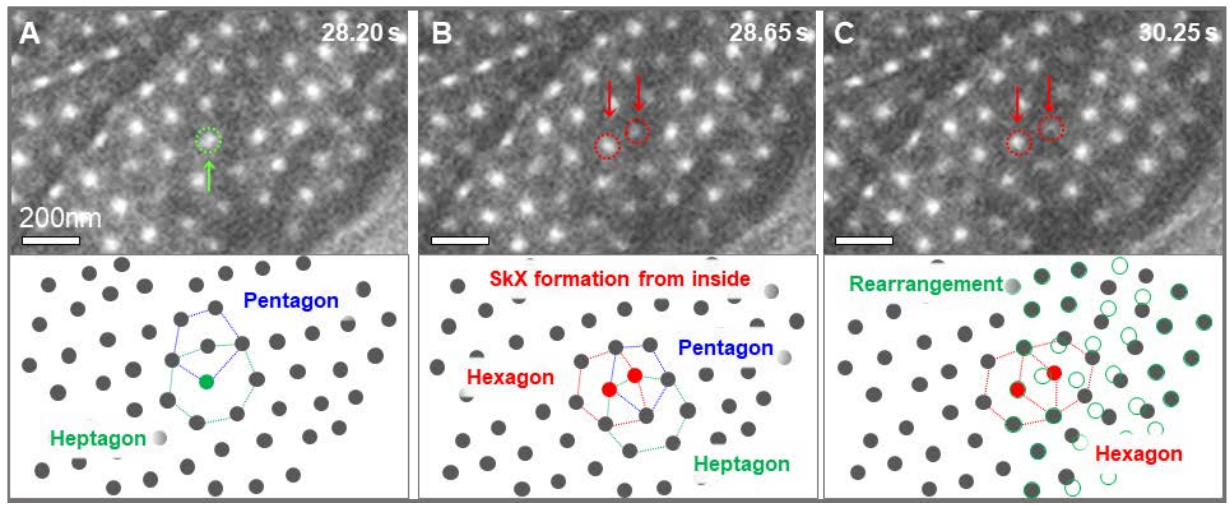

**Fig. 3 Self-splitting mechanisms of SkX growth**. Sequences of LTEM images from movie S5, taken at 88℃ under 100 mT, with schematic illustrations. Black circles indicate skyrmion positions in corresponding LTEM images. **(A)** Fivefold (pentagon) and sevenfold (heptagon) disclination inside the cluster. The green circle shows the center skyrmion of sevenfold disclination. **(B)** Splitting of the center skyrmion into two skyrmions (red). Hexagonal lattice (hexagon) was formed by splitting. The initial disclinations climbed closer to the surface of the cluster (the right side). **(C)** SkX lattice rearrangement to hexagonal lattice. Unfilled green circles are SkX lattice position of (B) before rearrangement and filled black circles with green outlines are after rearrangement.

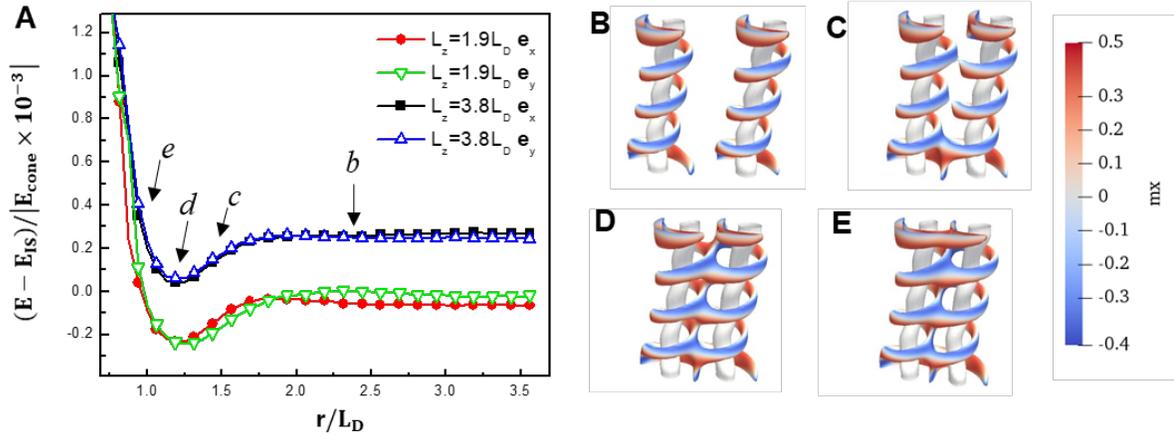

**Fig. 4. Interaction force and magnetic structure of two skyrmions**. (**A**) The energy density E as a function of inter-skyrmion distance **r** for separation in $x$ ($\mathbf{e}_x$) or $y$ ($\mathbf{e}_y$) direction. E is calculated with respect to the conical phase ($E_{cone}$) and single skyrmion ($E_{IS}$) energy under an external field (H) of $H/H_D = 0.5$, where $H_D$ is the saturation field. The E exhibits a decrease-increase Lennard-Jones like behavior with a minimal value at a distance slightly larger than the conical phase periodicity ($L_D$). (**B-E**) The magnetic structure of two interacting skyrmions marked by *b-e* in (A). The white and color streamlines in (B-E) are surfaces with z component of the magnetization $m_z = -0.4M_s$ and $m_z = 0.9M_s$, respectively, where $M_s$ is the saturation magnetization. The color indicates x component of the magnetization $m_x$.

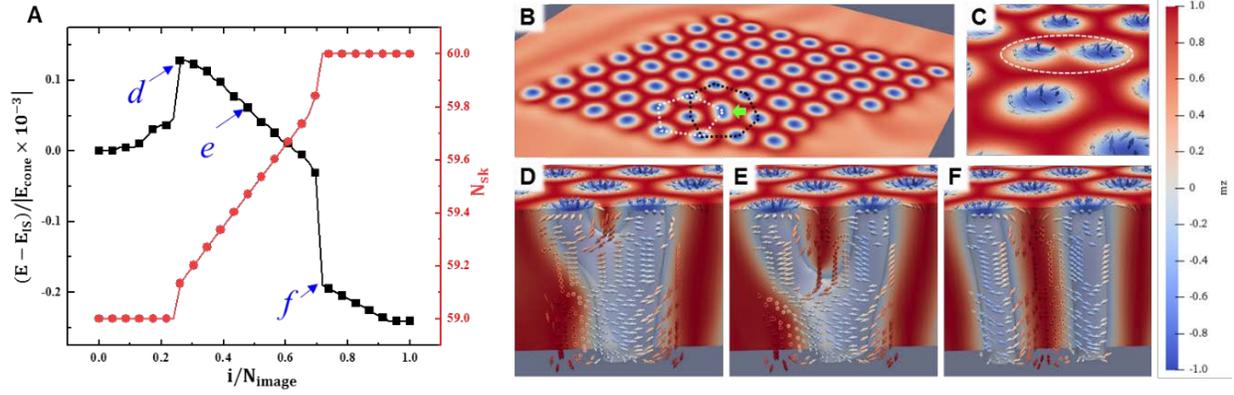

**Fig. 5**. **Energy profile and magnetic configuration during self-splitting of a skyrmion at 5-7 defect**. **(A)** The energy density E (left coordinate) and skyrmion number ($N_{sk}$) as a function of image number. E is calculated with respect to the conical phase ($E_{cone}$) and initial state ($E_0$) energy under an external field (H) of $H/H_D = 0.5$, where $H_D$ is the saturation field. $N_{sk}$ is averaged along z direction. **(B)** Top view of the SkX with a 5-7 defect used in calculation. The self-splitting skyrmion is marked by green arrow. Fivefold (white line) and sevenfold (black line) disclination is formed similar to experiment. **(C)** 2D view of magnetic structure of the splitting skyrmion (marked by white circle) with a magnetic monopole in the middle. **(D-F)** 3D view of magnetic structure of a self-splitting skyrmion corresponding to images marked by *d-f* in (A), respectively. The color plots in (B-F) indicate the z component of the magnetization $m_z$; arrows in (C-F) are direction of spins; Streamline in (D-F) is a surface with $m_z=-0.2$.

# Supporting information

# Mechanisms of Skyrmion and Skyrmion Crystal Formation from the Conical Phase

Tae-Hoon Kim,[1]† Haijun Zhao,[1,2]† Ben Xu,[3] Brandt A. Jensen,[1] Alexander H. King,[1,4] Matthew J. Kramer,[1] Cewen Nan,[3] Liqin Ke,[1]* Lin Zhou[1]*

[1] Ames Laboratory, U.S. Department of Energy, Ames, Iowa 50011, USA

[2] School of Physics, Southeast University, Nanjing 211189, China

[3] School of Materials Science and Engineering, State Key Laboratory of New Ceramics and Fine Processing, Tsinghua University, Beijing 100084, China

[4] Department of Materials Science and Engineering, Iowa State University, Ames, IA 50011, USA

*Correspondence to: liqinke@ameslab.gov; linzhou@ameslab.gov

† These authors contributed equally to this work.

# Section 1. Supplementary figures

## 1.1 Magnetic phase diagram of $Co_8Zn_{10}Mn_2$ thin plate

We systematically examined the temperature (T) and external magnetic field (B) dependence of different thermal-equilibrium states in a ~200 nm (110) $Co_8Zn_{10}Mn_2$ thin plate (fig. S2). Helical (fig. S1A), conical (fig. S1C) and SkX (fig. S1B) phases were observed, similar to literature[1]. LTEM images and corresponding fast Fourier transform (FFT) shows that the helical phase has stripe morphology with two-fold symmetry, whereas the SkX forms a hexagonal pattern with six-fold symmetry. Phase imaging analysis based on the intensity transport equation confirms that the skyrmion has a vortex-like swirling texture with a clockwise helicity[2]. No contrast is observed for the conical phase in the Lorentz image due to the q ∥ B ∥ [110] geometry, where q is the propagation vector of conical phase[1]. A phase diagram with a contour plot of skyrmion phase fraction based on LTEM observation is shown in fig. S1D, which demonstrates that SkX can be formed from the conical state by increasing temperature above 80 °C with an external field higher than 100 mT.

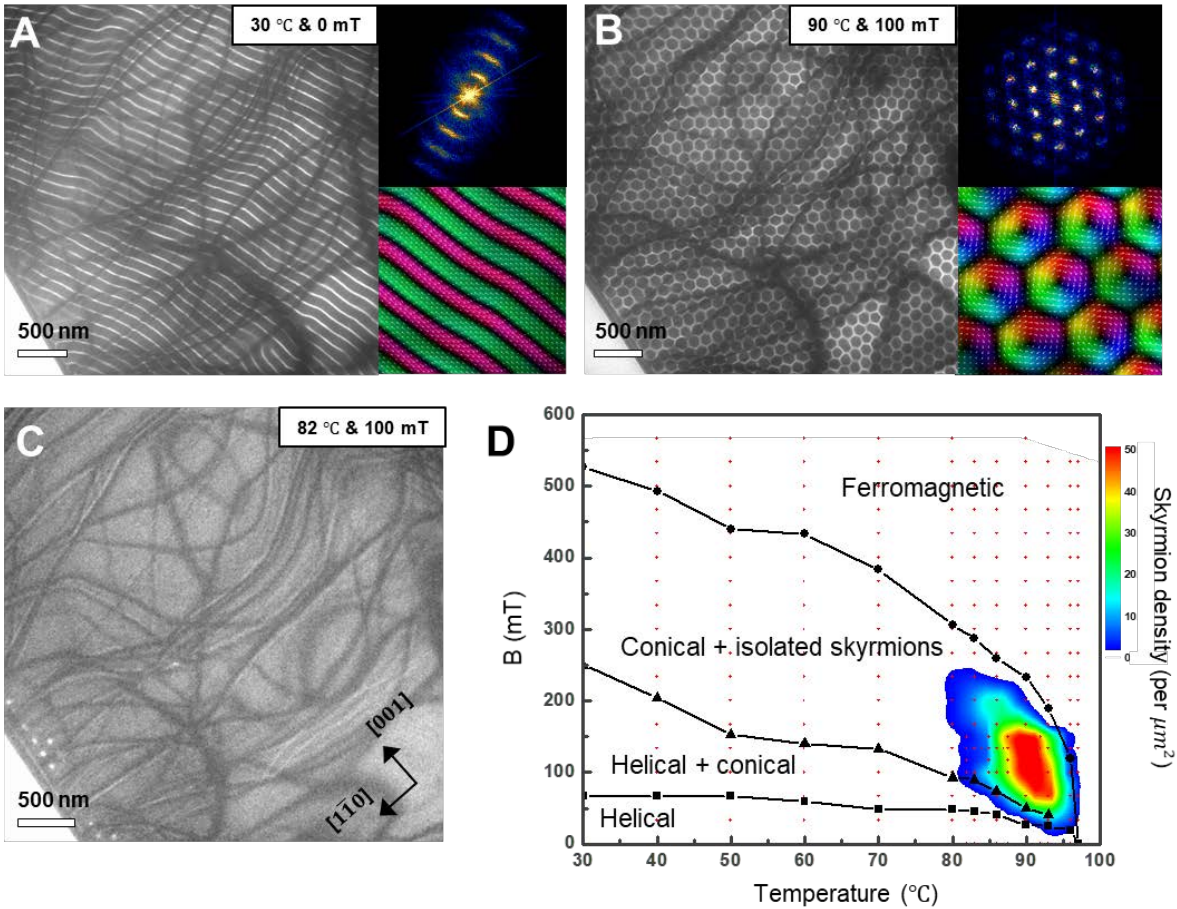

**Fig. S1 Magnetic structures and phase diagram in ~200 nm (110) $Co_8Zn_{10}Mn_2$ thin plate.** (**A**) Overfocused LTEM images of the helical phase under zero field at room temperature. Upper inset: corresponding reciprocal-space pattern. Lower inset: magnified in-plane magnetization map and white arrows inside indicate magnetization direction. (**B**) Hexagonal SkX lattice phase under 100 mT at 100℃. (**C**) The conical phase with a few isolated skyrmions under 100 mT at 82℃. (**D**) A contour plot of skyrmion density in the magnetic field ($B$) – the temperature ($T$) plane as deduced from LTEM observations. The white region has boundaries among the helical, conical and ferromagnetic phases. The color scale indicates skyrmion density per square micrometer. Red dots denote the experimental points.

**1.2 Thickness measurement of $Co_8Zn_{10}Mn_2$ using electron energy loss spectroscopy (EELS)**

The $Co_8Zn_{10}Mn_2$ (110) plate was fabricated using a focused ion beam system. The shape and thickness map are displayed in fig. S2.1, showing that the thin plate has a uniform thickness of ~200 nm.

We studied two other samples with a thickness of 140 nm and 240 nm, respectively (fig. S2.2). Both samples showed a similar growth mechanism as described in the main text.

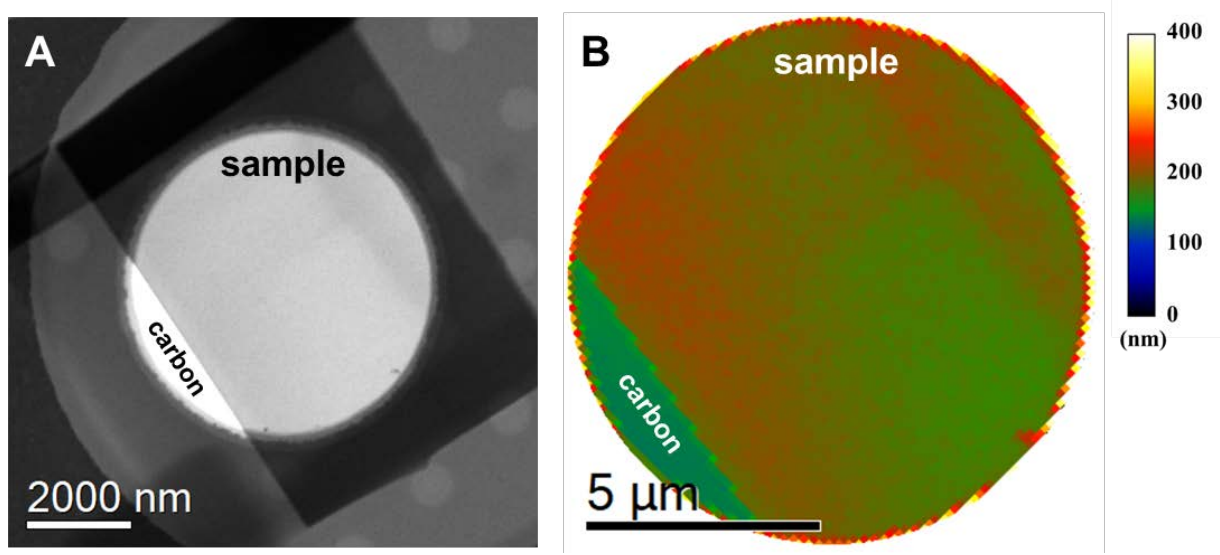

**Fig. S2.1 (A)** Bright field scanning transmission electron microscope image shows a TEM sample in the MEMS chip heater. **(B)** Thickness map obtained by EELS for a (110) $Co_8Zn_{10}Mn_2$ plate. The color bar shows the thickness scale.

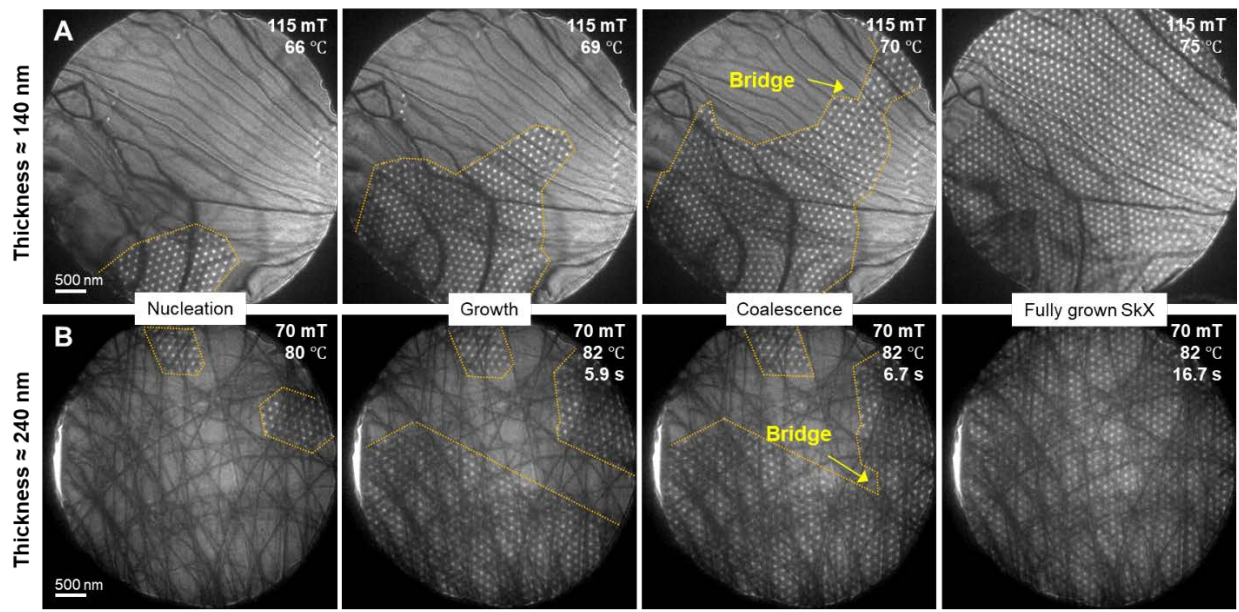

**Fig. S2.2** Sequential LTEM images of skyrmion growth obtained in the samples with a thickness of 140 nm **(A)** and 240 nm **(B)**, respectively.

## 1.3 LTEM image contrast evolution during skyrmion nucleation

Similar to the first skyrmion nucleation from skyrmion embryo as shown in Fig. 1C-D, contrast evolution was captured for the third and fourth skyrmions. Two skyrmion embryos (labeled as *c* and *d*) were formed at 5.85 s. Intensity profiles along the red line (fig. S3A) at each frame from 5.85 s to 6.75 s are shown in fig. S3G. The first mature skyrmion *a* shows almost constant intensity, whereas skyrmion embryo *c* kept about 60% of the maximum contrast of the mature skyrmion (*a*) for 0.35 s. Then, an instant contrast increase within 0.1 s into a mature skyrmion was observed. The normalized peak intensities of skyrmions *a* and *c* are shown in fig. S3H. Similar contrast evolution was observed in skyrmion embryo *d* and mature skyrmion *b*, but the intensity change was slightly interfered by diffraction contrast (a vague dark line across skyrmion embryo *d*). These results support that skyrmions grow through an intermediate metastable state, i.e. bobber or toron.

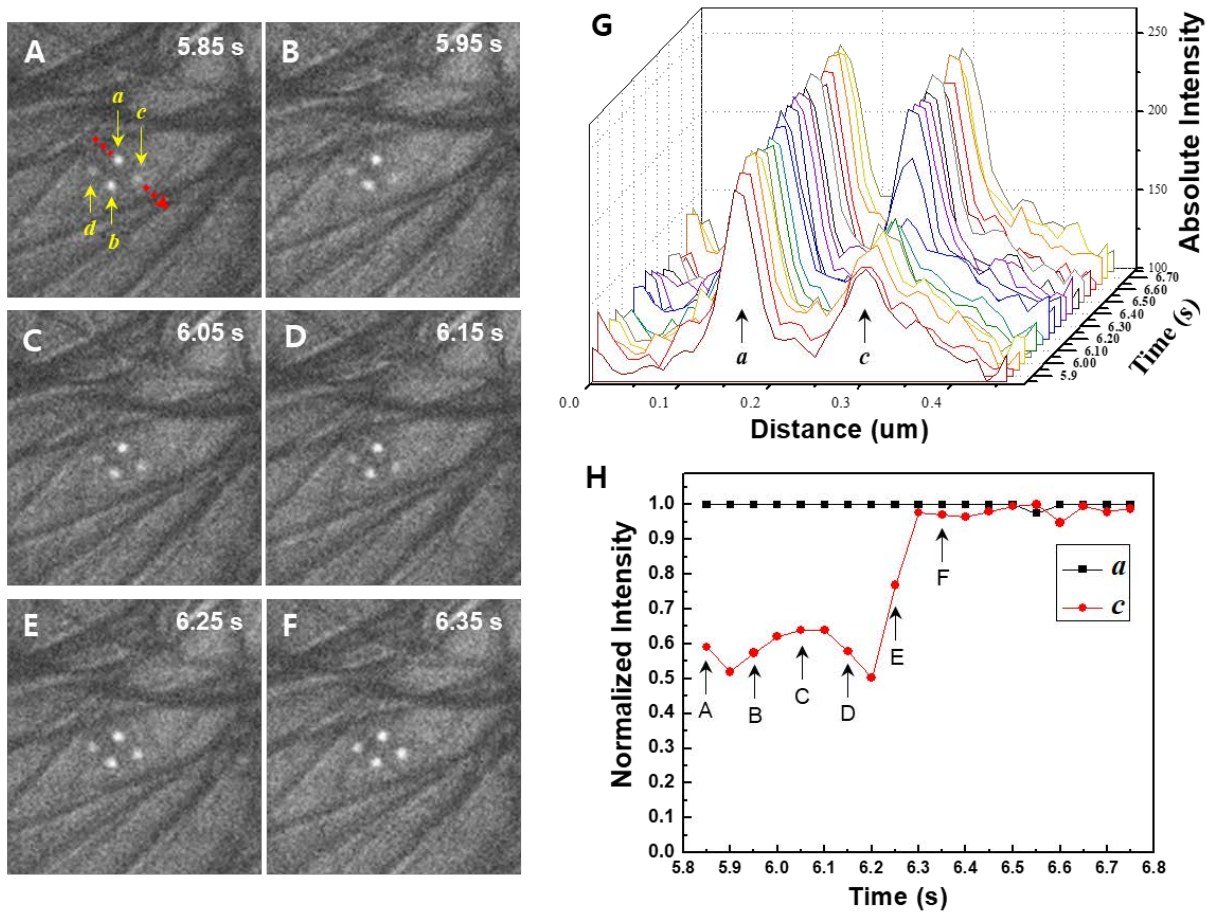

**Fig. S3** Time-resolved LTEM image showing contrast evolution during skyrmion nucleation. (**A-F**) The sequence of LTEM images from movie S1. Time indicates the time progressed after the sample reached 84°C under 135 mT. (**G**) Intensity profiles of skyrmions *a* and *c* obtained from sequential images from 5.85 s to 6.75 s. (**H**) The plot of the normalized peak intensity of skyrmions *a* and *c* versus time. Arrows with letters correspond to LTEM images of A-F.

## 1.4 Coalescence of skyrmion clusters: Oriented attachment (OA) with reversed sequence

While two neighboring clusters merge together, the smaller cluster tends to jump into the larger cluster as shown in movie S2. Fig. S4 shows misorientation reduction after the attachment.

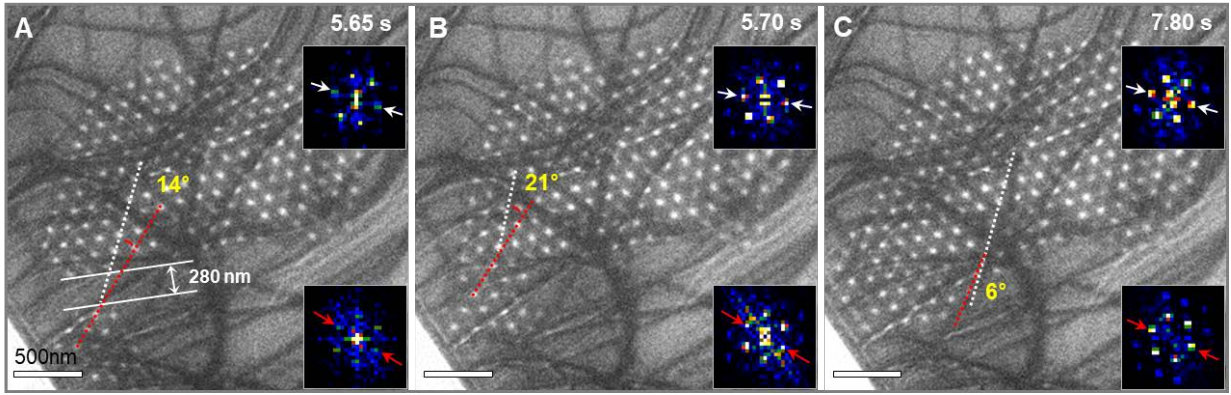

**Fig. S4** Sequence of LTEM images from movie S2, taken at 88°C under 100 mT. **(A)** Two clusters are ~280 nm apart right before merging. **(B)** Two clusters were attached in 0.05 s with an initial misorientation angle of 21°. **(C)** SkX rearrangement and misorientation angle was reduced to 6°.

## 1.5 Monomer-by-monomer addition (MA) of skyrmion

SkX lattice defects are formed at the surface of the clusters during SkX growth. The surface kinks with fivefold coordination act as preferred sites for MA skyrmion growth. As shown in fig. S5, isolated skyrmions grew at the position of fivefold disclinations to form a stable hexagonal lattice, creating another fivefold disclination just beside it at the same time. This MA process was repeated during SkX growth (movie S4).

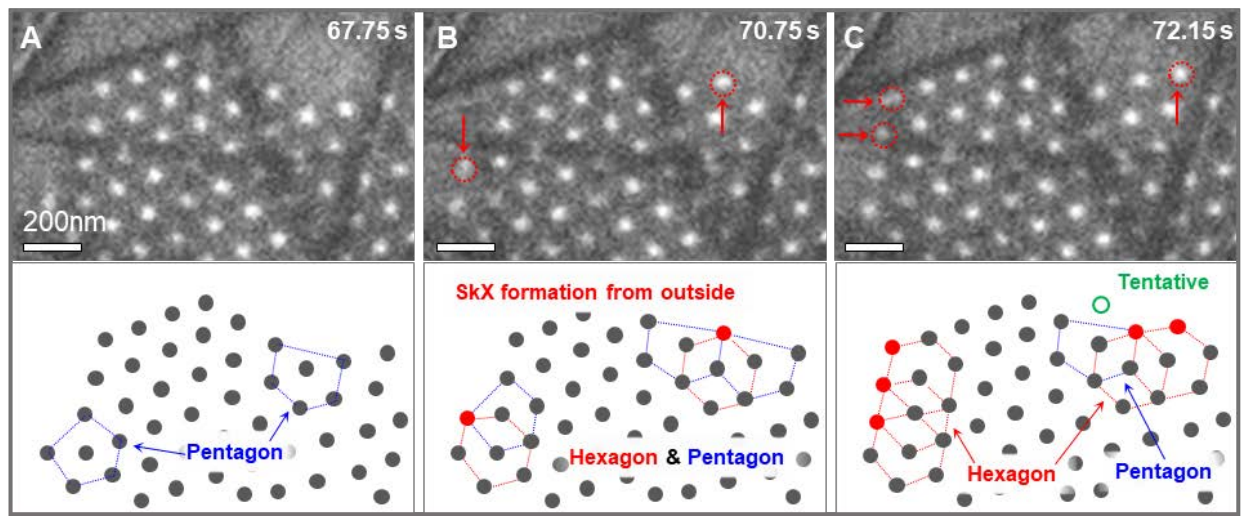

**Fig. S5 Monomer-by-monomer addition (MA) mechanism of SkX growth.** Sequences of LTEM images from movie S4, taken at 88°C under 100 mT, with schematic illustrations. Black dots in schematics indicate skyrmion positions of corresponding LTEM images. **(A)** Fivefold disclinations (pentagon) at the edge of the cluster. **(B)** Skyrmion attachment (red). Previous fivefold disclinations were changed to a hexagonal lattice (hexagon) and another fivefold disclination was formed just beside. **(C)** Continuous skyrmion attachment.

**1.6 Structural relaxation of SkX**

After the conical phase was totally replaced by SkX, the splitting of the FFT spots was still visible (fig. S6A). Domains with different orientations right after the phase transition were shown in the superimposed inversed FFT image (fig. S6B). This orientation difference disappeared with lattice oscillation and relaxation (fig. S6C, movie S6). Higher magnetic field and temperature were favorable to reduce disorder, similar to the previous report[3].

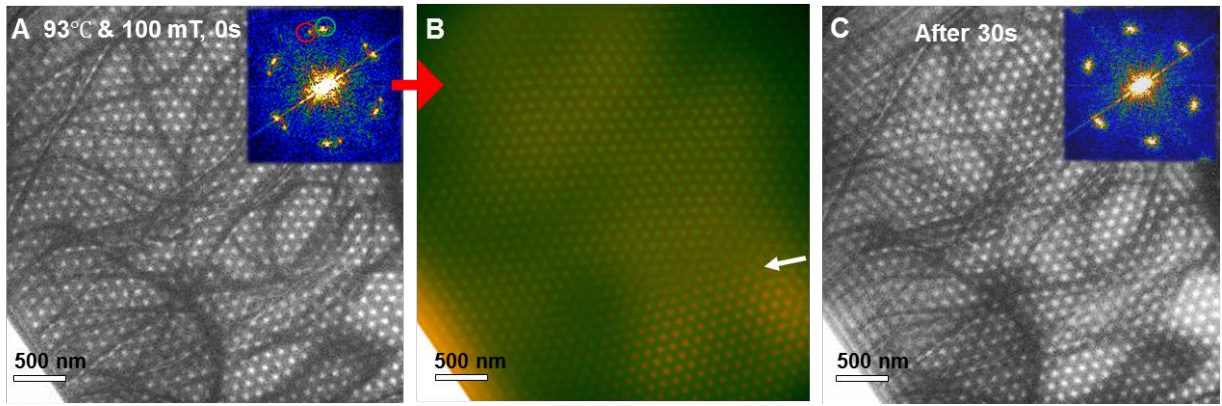

**Fig. S6 Structural relaxation of SkX (A)** SkX with different domains evidenced by the splitting of the Bragg peaks (inset FFT) right after phase transition completed at 90°C under 100 mT. **(B)** Superimposed inversed FFT image obtained using two inversed FFT images from different spots in the inset FFT of (A). Inversed FFT images from the red and green spots are indicated in red and green colors, respectively. Misoriented domain with a red color lattice is seen at the bottom right. The white arrow indicates the position of the grain boundary. **(C)** Single oriented SkX after 30 s at 90°C under 100 mT, and the inset FFT shows clear six-fold spots without splitting.

**Supporting videos**

Movie S1. Skyrmion and SkX nucleation

Movie S2. Oriented attachment (OA) during coalescence of SkX

Movie S3. Bridge formation during coalescence of SkX

Movie S4. Monomer-by-monomer addition growth of SkX

Movie S5. Internal self-splitting growth of SkX

Movie S6. Structural relaxation of SkX

## Section 2. Simulation details

### 2.1 Micro magnetic simulation

We use the standard model for magnetic states, for which the energy density functional is:

$$E = E_{ex} + E_{DM} + E_{Zeeman} = A(\mathbf{grad}\, \mathbf{M})^2 + D\mathbf{M} \cdot (\nabla \times \mathbf{M}) - \mu_0 \mathbf{M} \cdot \mathbf{H} \quad (1)$$

Here, $E_{ex}$, $E_{DM}$, $E_{Zeeman}$ are exchange energy, Dzyaloshinskii-Moriya (DM) coupling energy, and Zeeman energy, respectively. $\mathbf{M} = M_s(m_x, m_y, m_z)$ is the spin vector; $M_s$ is the saturation magnetization; $A$ is the exchange stiffness constant; $D$ is DM coupling constant; $\mu_0$ and $\mathbf{H}$ are permeability of vacuum and applied magnetic field, respectively. To calculate stable states as well as transition states, we numerically solve the Landau-Lifshitz-Gilbert (LLG) equation:

$$\frac{d\mathbf{m}}{d\tau} = -\mathbf{m} \times \mathbf{h}_{eff} - \alpha \mathbf{m} \times (\mathbf{m} \times \mathbf{h}_{eff}). \quad (2)$$

Here, $\mathbf{m} = \mathbf{M}/M_s$, $\tau$ is the effective time, $\alpha$ is the damping constant, and $\mathbf{h}_{eff}$ is the dimensionless effective field that defined as

$$\mathbf{h}_{eff} = \mathbf{H}_{eff}/M_s = (\partial E/\partial \mathbf{m})/(\mu_0 M_s^2).$$

Eq. (1) is solved by the Dormand-Prince method, which calculates fifth-order accurate solutions a fourth-order error estimate. The external field is applied along the z direction, i.e., $\mathbf{H} = (0,0,H)$. The cone states are given by:

$$\mathbf{m} = (\sqrt{1 - m_z^2}\cos\phi, \sqrt{1 - m_z^2}\sin\phi, m_z),$$

where $m_z = H/H_D$, $H_D = D^2/(2\mu_0 A M_s)$ is the saturation field; $\phi = 2\pi z/L_D + \phi_0$, $z$ is the z-direction coordinate, $L_D = 4\pi A/|D|$ is the helix period. $\phi_0$ is the initial phase of $z = 0$ layers. In our simulation, we set $\alpha = 0.5$ and $\phi_0 = 0$. Letting $A = L_D^2 * H_D * M_s/(8\pi^2)$ and $D = L_D * H_D * M_s/(2\pi)$, Eq. (1) becomes dimensionless, if we measure length in unit of $L_D$, external field in unit of $H_D$, and energy density in unit of $\mu_0 M_s H_D$.

We use a cubic distribution of spin sites with distance between neighboring sites $d = L_D/16$. To guarantee that $d$ is sufficiently small, we also calculate with $d = L_D/64$. Similar results were obtained. Periodic boundary conditions are used in the $x$ and $y$ directions and a free boundary condition is used in the $z$ direction. The size of the simulation cell is $128 \times 128 \times N_z$, where $N_z = L_z/d$ depends on the thickness of the sample. We set $N_z = 30$ to match with our experiment.

## 2.2 String method

We use a simplified string method[4] to investigate the skyrmion-skyrmion interaction and the self-splitting mechanism. Similar to elastic band methods, string methods are very useful to find minimum energy pathways (MEPs) for barrier-crossing transitions between two stable states, e.g., two stable spin configurations in our case. The transition pathway $Q$ is discretized into a set of images. Each image corresponds to a spin configuration of the system. To find the MEP, we start with an initial pathway $Q$ and calculate the energy $E$ of each image along $Q$, and evolve $Q$ to satisfy the following criterion

$$(\nabla E)_{\perp Q} = 0,$$

which means that the component of $\nabla E$ normal to $Q$ is zero, i.e., no tangential forces act on curve $Q$.

Starting from an initial curve, we iterate the following two steps to find MEP: (1) The evolution of every image with a short constant time step $\Delta t$ by LLG solver. (2) Interpolation /Reparametrization of images by equal-arc length. The path lengths of the transition steps are characterized by

$$\ell_a(i) = \sum_1^i \ell_a^{i,i-1} = \sum_1^i \sqrt{(m_{a,x}^i - m_{a,x}^{i-1})^2 + (m_{a,y}^i - m_y^{i-1})^2 + (m_{a,z}^i - m_{a,z}^{i-1})^2}.$$

Here, $\mathbf{m}_a^i = (m_{a,x}^i, m_{a,y}^i, m_{a,z}^i)$ is the spin vector of site $a$ of the $i$th image. Images $i = 0$ and $i = N$ correspond to the initial and final states, respectively, while the intermediate images represent the transition states of interest. After mapping the set of images $m^i$ to $\ell_{\{a\}}(i)$, we use them to interpolate a new set of intermediate images $m^i$ on a uniform mesh by setting the path length $\ell_a(i) = i * \ell_a(N)/N$. Once the new images are constructed, we renormalize them and repeat the step (1) to evolve each image by the LLG solver. We iterate the two steps (time evolution step and interpolation step) until convergence, more specifically, the change of energy corresponding to each image is sufficiently small between iterations. The number of images to represent the transition path, $N$, is kept the same during the simulation.

An educated initial transition path is required to start the calculation. If the initial path is too far from the real transition path, one may find a local minimum instead of the global minimum. In the present work, possible pathways are relatively limited, and we try various initial paths that we can imagine. Then, we compare the final energy profiles and identify the one with the lowest energy barrier as the most probable path. Moreover, a suitable imaginary path is also useful, as it helps to calculate many important features such as skyrmion-skyrmion interaction. In the following section, by constructing suitable initial pathways, we calculate skyrmion-skyrmion interaction and analyze the self-splitting process discovered in our experiment.

**2.3 Skyrmion-skyrmion interaction**

To calculate the skyrmion-skyrmion interaction, we start with a pair of well-separated skyrmions and gradually move them towards each other. During this process, skyrmions first bond to each other, then overlap and eventually merge into a single skyrmion. Thus, we construct a spin configuration of two well-separated skyrmions and a single-skyrmion state and use them as the initial and final states, respectively. The images of the initial transition path are constructed using a linear interpolation between the initial and final states:

$$\boldsymbol{m}_{\text{image}}^{(i)} = \{\boldsymbol{m}_S(\boldsymbol{R}_0) + \boldsymbol{m}_S(\boldsymbol{R}_0 + i \cdot d\boldsymbol{R}) - \boldsymbol{m}_{\text{cone}}\}.$$

Here, $\boldsymbol{m}_{\text{image}}^{(i)}$ is the configuration of the $i$th image, $\boldsymbol{m}_S(\boldsymbol{R})$ is a configuration with a single skyrmion located at $\boldsymbol{R}$, $d\boldsymbol{R} = d\hat{\boldsymbol{e}}$ is a vector position step we used to move the skyrmion with a distance $d$ along the $\hat{\boldsymbol{e}}$ direction. $\boldsymbol{m}_{\text{cone}}$ is the cone configuration before we include skyrmion. '{}' denotes the normalization of each spin. Starting from this initial path, we perform the loop of string method. During each step of the loop, skyrmions are first relaxed by solving LLG equation using a small time step. Then the interpolation process of the second step moves skyrmions back to their initial separation distance. The loop is continued until the curve of energy versus $|r| = i \times d$ converges (or no observable change of images for each step). As shown in Fig. S7, the calculated energy profile including two parts: The interacting part with $N_{sk}=2$, and the merging part with $N_{sk}<2$.

## 2.4 Self-splitting process.

We use the string method to investigate the intriguing transition path of skyrmion self-splitting (SS) at 5-7 lattice defect. We demonstrate that the SS process can be favorable in comparison to the nucleation and growth (NG) of a skyrmion at the defect.

For both SS and NG, the transition path can be classified into three possibilities: **A**. transition starts from the film surface, **B**. transition starts from the midsection of the skyrmion, and **C**. transition begins from the surface and inside simultaneously. Different initial pathways were constructed to simulate these three possible splitting mechanisms. We found that path **C** has a much higher energy barrier than that of **B** and **C** for thin foil thicker than $L_d$, consistent with the previous study[5,6]. Moreover, path **A** is more energetically favorable than path **B**. This is because the latter process requires generating two high energy transition point, instead of one (i.e., splitting point for SS and Bloch point for NG) during growth[7]. Thus, for both NG and SS, we consider the pathway that starts from the surface.

Now we identify the energetically favorable mechanism of growing a new skyrmion at the SkX defect from SS and NG pathway. For comparison, we also investigate the case of growing the second skyrmion next to an isolated skyrmion (IS). For both cases, we construct the same initial growth pathway by gradually rotating each spin from its first configuration (defect or IS) to the target arrangement (two skyrmions), but the rotation is triggered layer by layer with a small delay between adjacent layers starting from surface. Then we let the pathways evolve within the string method to find the MEPs. Interestingly, similar initial pathways evolve and end up with different skyrmion growth mechanisms — a NG-type MEP when a skyrmion grows next to an IS, but SS-type MEP when it grows at the dislocation site. Corresponding energy profiles are shown in Fig. 5A. The defect, where the splitting skyrmion is located, is shown in Fig. 5B. Fivefold and sevenfold disclination are similar to the experiment (Fig. 3A). To affirm that the growth mechanisms we found are indeed the MEPs, we try to start with different initial pathways to explore other possibilities, e.g., NG for the dislocation case and SS for the IS case. As shown in fig. S8A, SS for the IS case has a higher energy profile than that of NG. However, for growing new skyrmion at a dislocation site, the initial NG process ends up with a SS process, which clearly proves SS has a lower energy barrier. Furthermore, as shown in fig. S8, the energy barrier of creating a new skyrmion at a SkX defect is only ~25% of the energy barrier for creating one next to an isolated skyrmion. This is because the new skyrmion created at defect can bond to six neighbors instead of one. The magnetic structures of representative images are shown in fig. S8B-C. More details are presented in Movie S7-9.

The competition of NG and SS mechanism is less sensitive to film thickness since the energy difference between them is only from the small region close to the splitting, or Bloch point. However, it is sensitive to the external field, because SS requires elongation of existing skyrmion core, which is easy when the field strength is close to the critical value of phase transition from SkX to the helical phase. Finally, it is worth it to note that growing a new skyrmion on the defect is not the only way to remove the defect. We found that the dislocation can also gradually migrate to and annihilate at the boundary of the skyrmion cluster.

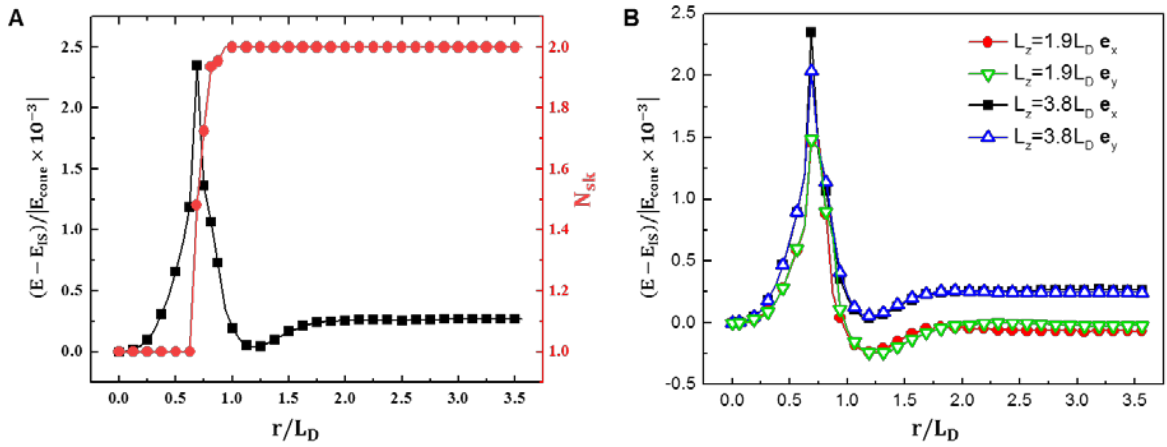

**Fig. S7. (a) The energy profile and skyrmion number $N_{sk}$ as a function of distance r calculated by string method for $L_z=3.8L_D$ and separation in x-direction.** Two insolated skyrmions move closer and merge into a single skyrmion. **(b)** The whole calculated curves for Fig. 4A.

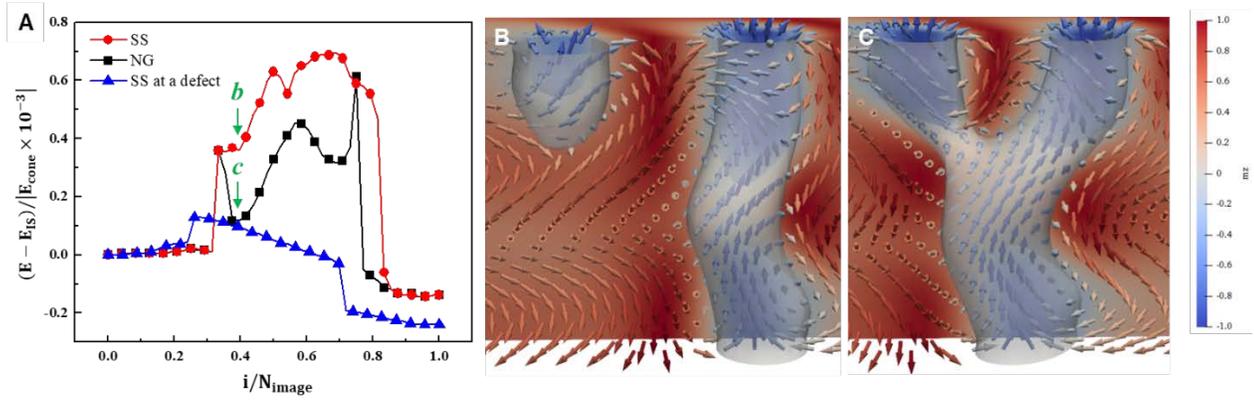

**Fig. S8**. **Energy profile and the pathway of creating new skyrmion**. **(A)** Energy profile of NG (red line, filled circles), SS (black line, filled squares) for creating a new skyrmion next to an isolated skyrmion, and SS (blue line, filled triangles) for splitting a skyrmion in a cluster located at a defect, respectively. **(B, C)** Magnetic structure of images marked by *a* and *b* in (A). The color plots in (B, C) indicate the z component of the magnetization $m_z$; arrows in (B, C) are the direction of magnetization; Streamline in (B, C) is a surface with $m_z = -0.2$.

**Supporting videos**

Movie S7. 3D view of a self-splitting skyrmion at a dislocation.

Movie S8. 2D view of a self-splitting skyrmion at a dislocation in a particular layer.

Movie S9. Growing new skyrmion next to an isolated skyrmion.

**References**

1. Karube, K. *et al.* Robust metastable skyrmions and their triangular-square lattice structural transition in a high-temperature chiral magnet. *Nat. Mater.* **15**, 1237–1242 (2016).

2. Yu, X. *et al.* Aggregation and collapse dynamics of skyrmions in a non-equilibrium state. *Nat. Phys.* 1–5 (2018).

3. Rajeswari, J. *et al.* Filming the formation and fluctuation of Skyrmion domains by cryo-Lorentz Transmission Electron Microscopy. **112**, 14212–14217 (2015).

4. E, W., Ren, W. & Vanden-Eijnden, E. Simplified and improved string method for computing the minimum energy paths in barrier-crossing events. *J. Chem. Phys.* **126**, (2007).

5. Leonov, A. O., Monchesky, T. L., Loudon, J. C. & Bogdanov, A. N. Three-dimensional chiral skyrmions with attractive interparticle interactions. *J. Phys. Condens. Matter* **28**, (2016).

6. Leonov, A. O. & Inoue, K. Homogeneous and heterogeneous nucleation of skyrmions in thin layers of cubic helimagnets. *Phys. Rev. B* **98**, 1–9 (2018).

7. Milde, P. *et al.* Unwinding of a skyrmion lattice by magnetic monopoles. *Science (80-. ).* **340**, 1076–1080 (2013).